\documentclass[aps,longbibliography,amsmath,amssymb,showpacs,preprint,twocolumn]{revtex4-1}
\usepackage[sort&compress]{natbib}

\usepackage{graphicx,tabularx}% Include figure files
\usepackage{dcolumn}% Align table columns on decimal point
\usepackage{bm}% bold math
\usepackage{color}

\usepackage{upgreek}

\renewcommand{\figurename}{Fig.~}

\begin{document}
\fontsize{11}{11} \selectfont
%\preprint{APS/123-QED}

%\title{Magnetotransport properties of a disordered
%graphite microwire produced by He$^+$ bombardment and embedded in a diamond crystal}

\title{Fabrication and electrical transport
properties of embedded graphite microwires in a diamond matrix}

\author{J. Barzola-Quiquia}
\email{Corresponding author: Tel.:+49 341 9732765,
\\E-mail: j.barzola@physik.uni-leipzig.de (Jose Barzola-Quiquia)}
\affiliation{Institute for Experimental Physics II,
University of Leipzig, 04103 Leipzig, Germany}
% \affiliation{Division of Thin Films Physics, Institute of Physics,
% \\Chemnitz University of Technology, 09107 Chemnitz, Germany}

\author{T. L\"{u}hmann}

\author{R. Wunderlich}

\author{M. Stiller}
%\email{m.stiller@studserv.uni-leipzig.de}

\author{M. Zoraghi}
\affiliation{Institute for Experimental Physics II,
University of Leipzig, 04103 Leipzig, Germany}

\author{J. Meijer}

\author{P. Esquinazi}
\affiliation{Institute for Experimental Physics II,
University of Leipzig, 04103 Leipzig, Germany}

\author{J. B{\"o}ttner  and I. Estrela-Lopis}
\affiliation{Institute of Medical Physics and Biophysics,
University of Leipzig, 04107 Leipzig, Germany}

\date{\today}
\begin{abstract}
  Micrometer width and nanometer thick wires with different shapes
  were produced $\approx 3~\upmu$m below the
    %
    %Amorphous microwires were produced $\approx 3~\upmu$m below the
  surface of a diamond crystal using a microbeam of He$^+$ ions with
  1.8~MeV energy.  Initial samples are amorphous and after annealing
  at $T\approx 1475$~K, the wires
    %of $\approx0.1~\upmu$m thickness
  crystallized into a graphite-like structures, according to confocal
  Raman spectroscopy measurements. The electrical resistivity at room
  temperature is only one order of magnitude larger than the in-plane
  resistivity of highly oriented pyrolytic bulk graphite and shows a
  small resistivity ratio
  ($\rho(2{\rm K})/\rho(315{\rm K}) \approx 1.275$). A small negative
  magnetoresistance below $T=200$~K was measured and can be well
  understood taking spin-dependent scattering processes into
  account. The used method provides the means to design and produce
  millimeter to micrometer sized conducting circuits with arbitrary
  shape embedded in a diamond matrix.

\end{abstract}
 %\verb+\pacs{#1}+
\maketitle

\section{Introduction}
\label{introduction}

Diamond is a natural allotrope of carbon, transparent, insulating and
the hardest natural material on earth.  Vavilov and
coworkers~\cite{VAV} have shown that it is possible to induce
graphitization in diamond by ion irradiation. One of the first works
trying to change the transport characteristics of diamond showed that
one can produce conducting regions by carbon implantation on the
diamond surface~\cite{HAUS1}. The value and temperature dependence of
the resistivity of ion implanted diamond layers was found to be
similar to that of amorphous carbon produced by
sputtering~\cite{HAUS2,HAUS1}.  In a recently published study,
micro-channels were fabricated in single-crystal diamond using a
microbeam of He$^+$ ions in the MeV energy range~\cite{pic12}. The
conductivity of the microchannels was improved substantially by
annealing treatment, achieving values similar to polycrystalline
graphite~\cite{pic12}.  The possibility to create a relatively long
conducting path of micrometer or even narrower width inside diamond is
interesting for possible electrical device applications where the main
electronic circuit remains well protected by the highly insulating and
biocompatible diamond matrix.

On the other hand, early studies
of 100~keV nitrogen and carbon implanted into nano diamonds, show
ferromagnetic hysteresis even at room temperature~\cite{TALAP}, and
recent studies on micrometer small areas on single crystal diamond
after irradiation with 2.2~MeV proton micro-beams provided hints on the
existence of magnetic order~\cite{mak14,mak16}. The authors found that for
fluencies below $8.4\times 10^{17} {\rm cm}^{-2}$, a very weak magnetic
response was observed at room temperature. In both mentioned
studies, the origin of the magnetism was related to the defects
produced by the irradiation.

In this work, a similar technique as in Ref.~\cite{pic12} was used to
produce conducting microwires beneath the diamond surface using He$^+$
irradiation. After two annealing steps, different degrees of
graphitization of the microwire were reached. Our aim was also to
check, whether those conducting structures can show some degree of
magnetic order. As was shown in several works in the past, defects
like vacancies and/or non-magnetic ions within the graphite structure
as well as in a large number of materials, can trigger magnetic order
even above room temperature, a phenomenon called defect-induced
magnetism (DIM)~\cite{dim13}.  Therefore, one can expect that a
defective graphitic structure, like the one we produce within the
diamond structure by ion irradiation and after annealing, may show
some magnetic response. The possibility of having conducting and
magnetic microwires within a pure diamond matrix provides a further
interesting option for future application.

\section{Experimental Details}
\label{procedure}

\subsection{Preparation of sub-micrometer width and millimeter long graphite wires}

To produce a graphite-like microwire (GLM) inside diamond, we used
a polished single crystal (100) diamond of the company
Element Six, with a nitrogen concentration $< 1$~ppm and a
boron concentration $< 0.05$~ppm. The diamond sample was produced
by chemical vapor deposition (CVD) with dimension $\rm
2.6\times2.6\times0.3~mm^3$.
%%%%%%%%

The He$^+$ ion irradiation used to
produce the wires in diamond was realized
using a very stable high-energy ion nanoprobe in the linear
accelerator LIPSION at the University of Leipzig.  The wires within the diamond structure was produced
with a microbeam of 1.8~MeV energy and an ion current of 2.4~nA.

\begin{figure}
\includegraphics[width=\columnwidth]{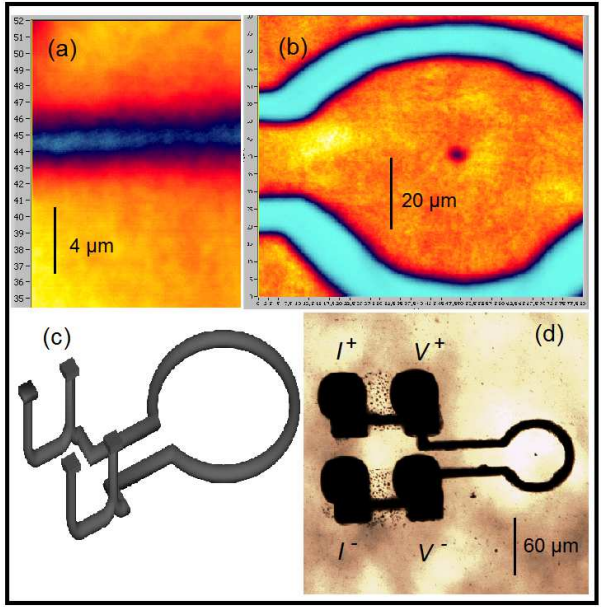}
\caption{\label{fig:fig5s} Image of graphitized structures. (a) Confocal image of a wire with a width in
the order of $\approx 1~\upmu$m. (b) Confocal image of a wire in shape of a loop. All images taken after annealing (c) 3D sketch of the loop (d) optical picture
of the wire loop showing the current-voltage ($I-V$) contacts at the surface of the diamond. }
\end{figure}

In our accelerator facilities, using the high-energy ion nanoprobe, we
can obtain a sub-micrometer ion beam diameter~\cite{SPEM} using a two
magnetic quadrupole double lens system. Also, to produce the long size
microwire inside the diamond, we need a high ion current and energy
stability. This two important conditions are fulfilled with the ion
accelerator Singletron$^{\rm TM}$ from the Dutch company High
Voltage Engineering Europa B.V.~\cite{MOUS}. The ability to produce
complex two dimensional structures (2D) is obtained with the help of a
raster unit, which is located between the quadrupole lenses and the
sample. Using a self-made program we can produce any desired 2D
structure by deflection of the ion beam. According to our experimental
conditions, we can continuously deflect the He$^+$ ion beam within a
total area of ($1780\times1780)~ \upmu \rm m^2$. As examples, using a
beam diameter of $\approx 1~\upmu$m, a microwire of similar width, see
\figurename~\ref{fig:fig5s}(a), inside a diamond substrate was
produced and after annealing, we have graphitized the wire. As a proof
of the ability of our ion nanoprobe to produce complex structures (see
also other structures in supplementary information), we have prepared
a graphitized loop, see \figurename~\ref{fig:fig5s}~(b-d), which can
be used to generate small magnetic fields and/or microwave fields in
order to manipulate, e.g., the spin states of NV centers in
diamond. The images in \figurename~\ref{fig:fig5s} were obtained with
a self made confocal optical microscopy implemented with a lens
(Olympus MPlanApo 100x/NA0.95), a laser beam with $\lambda = 532$~nm
and a lateral and depth resolution of $\approx 300$~nm and
$\approx 1~\upmu$m respectively.
Summarizing, we are able to produce long wires with a width of
$\approx 1~\upmu$m and desired shape. These results show the
possibilities to produce 2D graphitized structures inside diamond for
future applications.

%%
%The He$^+$ ion irradiation used to
%produce the microwire in the sample was realized
%using a very stable~\cite{MOUS} high-energy ion nanoprobe in the linear
%accelerator LIPSION at the University of Leipzig.  The three
%dimensional microwire within the diamond structure was produced
%with a microbeam of 1.8~MeV energy of $3~\upmu$m in diameter and
%with an ion current of 2.4~nA.

\subsection{Irradiation Effects on Diamond and Annealing Treatments}

After irradiation of diamond with He$^+$ ions of energy and fluence
similar to the ones we used in this work, an amorphous phase is
produced at the region where the irradiated ions stop~\cite{VAV}.
By annealing at high
temperatures, this amorphous carbon region can later be transformed into
graphite or back to diamond structure, which is correlated to the density of vacancies produced by the
irradiation. There exists an estimated critical vacancy density
 $N_{\rm v0} \sim 10^{22}\rm cm^{-3}$~\cite{UZAN} to induce
graphitization in the material after ion irradiation at room
temperature and after annealing. According to other experimental
work and supported by molecular dynamics simulation, the critical
density necessary to graphitize the surface~\cite{KAL} was
determined to $N_{\rm v0}\approx9\times10^{22}\rm cm^{-3}$. Raman
measurements indicate that after annealing above $T\approx 700$~K
the graphitization process begins by formation and growth of
$sp^2$ bonded nanoclusters~\cite{KAL}.

\begin{figure}
\includegraphics[width=\columnwidth]{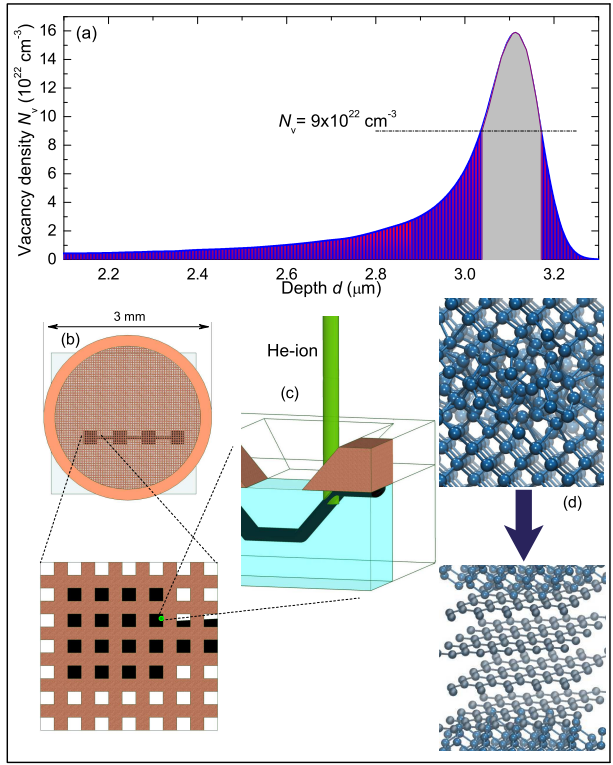}
\caption{\label{fig:fig1} Sketch of the sample and contact
  preparation. (a) SRIM simulation showing the vacancy density versus the
 penetration depth for the fluence and energy (see main text)
 used in this work. (b) In the first irradiation step, the
  microwire was produced without a
  mask, afterwards, the diamond substrate was covered using a copper grid.
    (c) The sketch shows how contacts from the embedded wire
  were grown on the substrate surface. (d) Sketch of the amorphous carbon region (top) and the region transformed into
graphite (bottom) by annealing at high temperatures.}
\end{figure}

We have irradiated the diamond sample with He$^+$ ions using a
fluence of $\phi = 5.5\times10^{17}\rm cm^{-2}$, which gives a
vacancy density $N_{\rm v} \gtrsim 9\times10^{22}\rm cm^{-3}$ in a
specific depth, see \figurename~\ref{fig:fig1}(a) estimated from
Monte-Carlo simulation done by SRIM~\cite{SRIM}.
The program simulates the atomic-displacement cascades in solids
on the base of the binary-collision approximation to construct
the ion trajectories~\cite{ROBIN}. For the calculations we have assumed a
dislocation energy of 52~eV~\cite{SAAD}. This vacancy density is
around the minimum value for graphitization after annealing of the
ion-produced amorphous carbon phase.

%
%With the microbeam we are able to produce wires of $\lessapprox 1
%\mu$m width and 1.8~mm length (see supplementary information).
To
facilitate the transport measurements performed in this study, the
produced GLM had dimension of $15~\upmu$m width, $1350~\upmu$m
total length and a estimated thickness of $130\pm5$~nm, see
\figurename~\ref{fig:fig1}(a). According to the SRIM calculations
the major damage density produced in the diamond sample is
located at $\approx 3.1~\upmu$m beneath the diamond surface, which
was verified by confocal micro-Raman spectroscopy (CRS), see
Section~\ref{raman}.

After He$^+$ irradiation, the sample was first annealed at $T
\approx1475$~K in a vacuum chamber with a pressure of $P \approx
5\times10^{-6}$~mbar (heating rate of 15~K/min and cooling rate of
10~K/min) for 4~hours for the first and for 2 additional hours in
the second annealing treatment.  After the first and also after
the second annealing process the sample was placed in an oxygen
plasma chamber at room temperature in order to remove the
conducting carbon-based thin layer formed at the surface after the
annealing, to avoid any contribution to the transport measurements.

\subsection{Electrical Contacts to the Microwire, Transport and Raman Measurements}

The contacts for the electrical measurements have  to be done at
the surface of the substrate. For this purpose we used a
commercial copper grid used for transmission electron microscopy
(2000 mesh), having the advantage  that the grid has a wedge shape
allowing a continuous change of the He$^+$ ions penetration depth
inside the diamond sample, see Fig.~\ref{fig:fig1}(b-d). At the
surface of the sample and at a distance of $\approx 450~\upmu$m,
square-like contacts with dimension of $\approx
50\times50~\upmu{\rm m}^2$ were prepared in direct electrical
contact with the embedded microwire, Fig.~\ref{fig:fig1}(c). A
similar but more complicated method was already used in
Ref.~\cite{pic12}. Afterwards, the electrodes were made by
sputtering of Cr/Au directly at the top of the square-like regions
at the surface, after an electron beam lithography process.
Finally, the contacts to the chip carrier, where the sample was
fixed for the transport measurements, were produced using silver
paste and gold wires.

Resistance measurements were carried out using the conventional
4-points method using an AC Bridge (Linear Research LR-700) in the
temperature range of 2~K to 310~K. For the current-voltage
(\textit{I--V}) measurements, we used the Keithley DC and AC
current source (Keithley 6221) and a nanovoltmeter (Keithley
2182).  The resistance and its magnetic field dependence were
measured with a commercial cryostat from Oxford Instruments with a
superconducting solenoid that provides a maximum field of
$\pm~8$~T perpendicular to the main axis of the microwire.  Raman
characterization was carried out at room temperature using a
confocal micro-Raman microscope (alpha300+, WITec company) with an
incident Laser light of $\lambda = 532$~nm, a lateral resolution
of $\approx 300$~nm and axial resolution of $\approx 900 $~nm.

\section{Results and Discussion}
\label{results}

\subsection{Raman results}
\label{raman}

\begin{figure}
\includegraphics[width=\columnwidth]{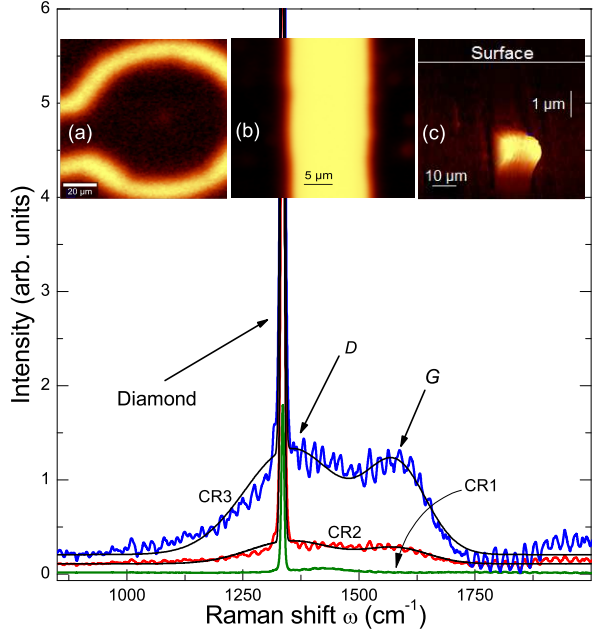}
\caption{\label{fig:fig6} Raman results measured at the surface and
  inside the diamond substrate.  The curves CR1 and CR3 were obtained
  fixing the focus of the Raman microscope at the surface (CR1) on the
  as-received sample, and in a depth of $3~\upmu$m (CR3) after
  irradiation and second annealing. The curve CR2 was obtained from
  the measurements at the surface after He$^+$ irradiation and the
  second heat treatment. The continuous lines through the experimental
  curves are the fits to the data. The insets show confocal Raman
  images, in (a) the loop shown in \figurename~\ref{fig:fig5s} can be
  seen, (b) the image shows a close up of a graphite wire and (c)
  shows an area perpendicular to the surface, i.e.~as function of the
  sample depth. The bright part represents the $G$-graphite peak. }
\end{figure}

To get information about the structural properties of the measured
wire produced inside the diamond sample, we have used CRS, which is probably the only method to get
information about the microstructure produced inside the diamond
without destroying it. \figurename~\ref{fig:fig6} shows
the Raman results.

The curve named CR1 in \figurename~\ref{fig:fig6} was measured at
the surface of the as-received diamond sample. The Raman peak at
$\approx 1332~\rm cm^{-1}$ is the characteristic peak for carbon
in the diamond structure. The small bump at $\approx 1430~\rm
cm^{-1}$ is related to the appearance of disorder at the surface
of the sample during the polish process~\cite{MA}.  The curve CR3
obtained after irradiation and the second annealing was measured
with the focus at $\approx 3~\upmu$m depth from the diamond
surface. The results  confirm the estimated depth  from the SRIM
calculations. The results of the curves CR2 and CR3 show three
characteristic peaks, one at $\approx 1332~\rm cm^{-1}$
corresponding to the diamond structure, a second peak due disorder
graphite, the so-called $D$ peak at $\approx 1350~\rm cm^{-1}$.
The third and the most important peak for the characterization of
graphite structure, called the $G$-peak, appears at $\approx
1580~{\rm cm}^{-1}$ as a consequence of the double degenerate zone
center $E_{2g}$ mode.

Our Raman results resemble those obtained in \cite{KAL}, specially
for similar annealing temperatures. The results CR2 and CR3 can be
very well fitted using Gaussian functions centered at the
aforementioned Raman peaks. The results are shown as continuous
black lines in~\figurename~\ref{fig:fig6} and describe very well
the experimental results. From the fits we get also information
about the peak intensity $I_G$ and $I_D$ corresponding to the $G$
and $D$ peak respectively.

From these results it is possible to estimate the crystal
size~\cite{CANC} $L_a$ using Eq.~(\ref{eq:La}):
\begin{equation}\label{eq:La}
L_a({\rm nm})=
(2.4\times10^{-10})\lambda_l^4\left(\frac{I_D}{I_G}\right)^{-1},
\end{equation}
which correlates the crystal size $L_a$ with the integrated intensities of the $D$ and
$G$ peaks and the laser excitation wavelength
$\lambda_l$~=~532~nm. Using this equation, we obtain $L_a=8\pm 2$~nm, similar
to the results of Rubanov~\textit{et.-al.} using transmission electron
microscopy~\cite{RUBA}, where diamond samples were irradiated
with He$^+$ ions using a fluence between
$3\times10^{16}-10\times10^{16}\rm{cm}^{-2}$ followed by an annealing
for 1~h at $1400~^\circ$C.

\subsection{Temperature Dependence of the Electrical Resistance}
After the first annealing the sample shows (at low temperatures)
non-linear \textit{I--V} curves, indicating that non graphitized
regions remain in the sample which act like barriers. It has been shown
that, whenever a barrier is present between the conducting grains, the
resistance $R(T)$ and magnetoresistance ($MR$) depend on the applied
current like in multi-wall carbon nanotubes
bundles~\cite{JBQMWCNT}. Here we do not discuss the $R(T)$ and \textit{I--V}
results after the first annealing (see supplementary information) because our interest lies in the
behavior of the transport properties in the Ohmic regime, without any
influence of potential barriers, this is obtained after the second
annealing treatment.

\begin{figure}
\includegraphics[width=\columnwidth]{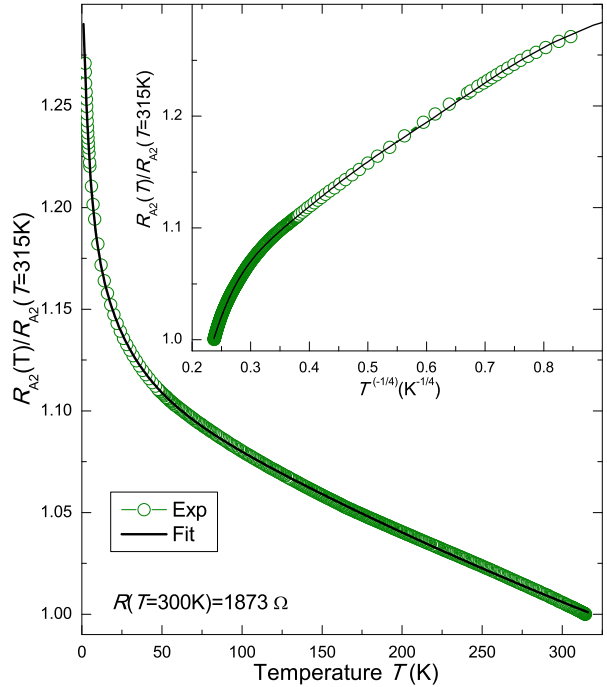}
\caption{\label{fig:fig3} Temperature dependence of the resistance
(after the second annealing). The inset shows the resistance
versus $T^{-1/4}$. The continuous lines are the fits results.}
\end{figure}

The resistance results after the second annealing are shown
in~\figurename~\ref{fig:fig3}. The experimental results from 2~K
to 315~K are shown as open symbols. The observed temperature
dependence is clearly different from the one we obtained after the
first annealing treatment.
%\subsubsection{After the second  annealing treatment}
The resistance ratio $R_{A2}(2{\rm K})/R_{A2}(315{\rm K}) \approx
1.275$ is one order of magnitude smaller than the one after the
first annealing, and of the same order as for nano-graphite
films~\cite{CHOL} and few layer graphene films~\cite{JB1,MZ1}.
The current-voltage curves after second annealing are linear in all
measured temperature range  and are shown in the supplementary information.

%the amorphous regions were completely crystallized after the second annealing.
%The linear \textit{I--V} characteristics and

The temperature dependence of the resistance, see
Fig.~\ref{fig:fig3}, indicates the existence of two different
regions, one below and the other above $T\approx 75~$K. We have
identified (see supplementary information) that the dominant
mechanism at high temperatures is the so-called Mott variable
range hopping (VRH)~\cite{MOTT}, given as:
\begin{equation}\label{eq:Mott}
R_{VRH}(T)= R_M\exp\left[\left(\frac{T_{0}}{T}
\right)^{1/4}\right],
\end{equation}
where $T_{0}$ is a characteristic temperature coefficient defined
as:
\begin{equation}\label{eq:Th}
T_{0}= \frac{18}{k_{\rm B}\xi^3 N(E_{\rm F})},
\end{equation}
and $\xi$ is the localization length, $N(E_{\rm F})$ the
density of states at the Fermi level and $R_M$ is a
constant prefactor. In order to fit the resistance over all the
measured temperature range, we need to include an extra transport
mechanism in parallel (we have also checked other configurations,
see supplementary information) given as:
\begin{equation}\label{eq:Rtot1}
R_{A2}(T)^{-1}=R_{VRH}(T)^{-1}+R_{m}(T)^{-1},
\end{equation}
with
\begin{equation}\label{eq:Rtot2}
R_{m}(T)=R_{0}+R_aT+R_{b}\cdot {\rm exp}(-E_{a}/k_{\rm B}T),
\end{equation}
a metallic-like contribution, which dominates at low temperatures.
The coefficients $R_0$, $R_a$, $R_b$ as well as the activation
energy $E_a$ are free parameters. $R_{0}$ a residual temperature
independent resistance that we include in the metallic-like
contribution only,  provides a saturation of the resistance at low
temperatures. The contribution $R_{m}(T)$ was already used to
explain the temperature dependence of the resistance in bulk
graphite~\cite{MATSU}, few layer graphene samples~\cite{GARC,MZ1}
and nano-graphite thin films~\cite{CHOL}. Its origin is related to certain
interfaces formed between the graphite crystallites. We can fit
the experimental resistance data to Eq.~(\ref{eq:Rtot1}) very
well. The results are shown as continuous lines
in~\figurename~\ref{fig:fig3} and the fitting parameters are
listed in Table~\ref{tab:Tab1}.
\begin{table}[t]
\begin{tabularx}{\columnwidth}{XXXX}
  \hline\hline
  $R_M(\Omega)$ &$T_{0,{\rm Mott}}({\rm K})$&$R_{0}(\Omega)$& $R_{a}(\Omega)$    \\
  \hline
  686.66      &  723.6      & 2431.31        &     20.08     \vspace{0.5cm}\\
  $R_{b}(\Omega)$&$E_{a}({\rm meV})$&$N(E_F)$~(eV$^{-1}{\rm cm}^{-1})$\\\hline
  276.4          &     1.96        & $2.3 \times 10^{24}$         \\
  \hline\hline
\end{tabularx}
\caption{Summary of parameters obtained after fitting the resistance $R_{A2}(T)$ using Eq.~(\ref{eq:Rtot1}). }
\label{tab:Tab1}
\end{table}

The inset of \figurename~\ref{fig:fig3} shows clearly the temperature
ranges where each transport mechanism dominates the electric
transport. We have assumed a localization length of $\xi\approx 0.5$~nm~\cite{TL-16}, which was
estimated for similar samples, by means of spectroscopic ellipsometry, Raman spectroscopy and transport
measurements. The localization length used in this work is close to the value of 1.2~nm used
by Hauser et-al~\cite{HAUS1}. We
estimate $N(E_F) \sim 2.3 \times 10^{24}$~eV$^{-3}$cm$^{-1}$, which is of
the same order as reported for graphitic materials~\cite{RAJ} and
multi-walled carbon nanotubes~\cite{KHA}. Values of $N(E_F)$ of
several orders of magnitude lower than our result were found for
similar ion irradiated diamond, but these samples were not
annealed~\cite{HAUS1,OLIV1} or even measured when a barrier was
present~\cite{OLIV1}. Finally, the calculated resistivity of the GLM
is $\rho_{ (300~{\rm K})}=7 \ldots 8 \times 10^{-5}~\Omega$m, one
order of magnitude larger than bulk graphite in-plane resistivity
($\rho_{ (300~{\rm K})}=0.5\times
10^{-5}~\Omega$m)~\cite{KEMP,JB1,MZ1} or the resistivity of few layer
graphene
($\rho_{ (300~{\rm K})}=3 \ldots 50\times 10^{-5}~\Omega$m)~\cite{JB1,MZ1}
and three orders of magnitude lower than pure amorphous carbon
($\rho_{ (300~{\rm K})} > 10^{-2}~\Omega$m)~\cite{GRILL,HAUS3}. However, the GLM resistivity is
comparable to other nano-graphite thin films prepared by chemical
vapor deposition and aerosol assisted chemical vapor deposition
($\rho_{ (300~{\rm K})}=3 \ldots 8\times 10^{-5}~\Omega$m)~\cite{CHOL}.  From
experiments it is known that the resistivity ratio between the
in-plane ($\rho_a$) and out-of-plane ($\rho_c$) resistivity in graphite
is in the order of $\rho_c/\rho_a\approx10^3-10^4$~\cite{KEMP,KOPE1},
indicating that the transport in our sample is  dominated by the
in-plane resistivity. Further, it means that the produced GLM have
nano-crystals with a preferential \textit{c}-axis normal to the
substrate surface. The obtained transport characteristics of the
produced GLM are important for future applications, because the
resistivity and its temperature dependence make this GLM interesting
to be used for electronic circuits in a broad temperature range.

\subsection{Magnetoresistance}

\begin{figure}
\includegraphics[width=\columnwidth]{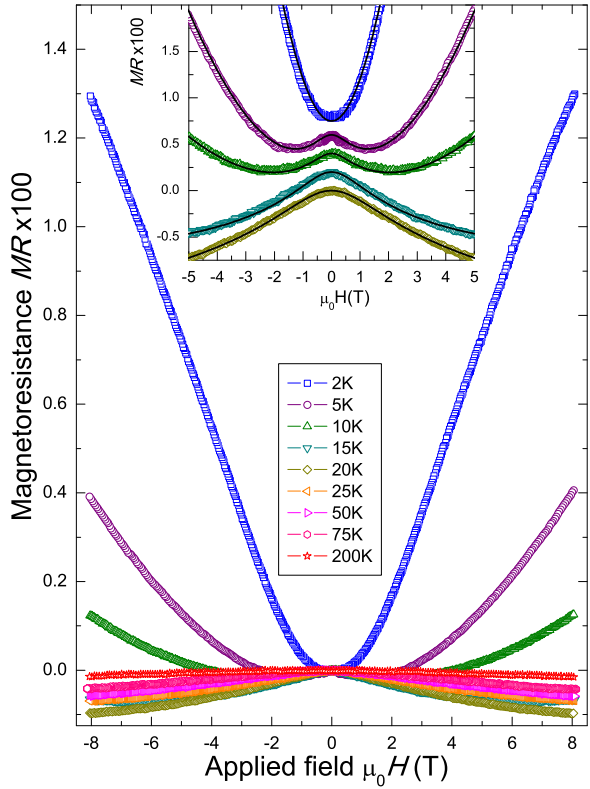}
\caption{\label{fig:fig5} Magnetoresistance at
  different constant temperatures. The inset shows the $MR$ obtained at $T \eqslantless 20~$K.
  The lines through the points are  fits to Eq.~(\ref{eq:MR}).
  The results were shifted by a
  constant in the vertical axis in order to facilitate a detailed view.}
\end{figure}

To obtain any information of the magnetic properties of the GLM
there are no experimental methods other than transport because the
wire is not only embedded inside the diamond matrix but also its
mass is too small to be measured with commercial magnetometers.
The magneto-transport measurements at different constant
temperatures were done with an external magnetic field applied
perpendicular to the current and main axis of the GLM. The results
of the magnetoresistance $MR$ defined as $MR=[R(B)-R(0)]/R(0)$
are shown in~\figurename~\ref{fig:fig5}.

We observe a magnetic field dependence of the resistance, which is
positive at $T=2$~K, and negative in the field range of -3~T~$< B
<$~3~T at 5~K and 10~K. At temperatures $T>10$~K the $MR$ is
negative in all the field range. The positive $MR$ at very low
temperatures can be understood as a consequence of the strong
Lorentz contribution. This tends to vanish at high temperatures.
The negative $MR$ can be attributed to a spin dependent scattering
process. Due to the relatively high temperatures and magnetic
field range where the negative $MR$ is observed, we can rule out weak
localization effects.  According to the theory developed by
Toyozawa~\cite{TOYO} and later modified by Khosla and
Fischer~\cite{K-F}, the $MR$ of systems with localized magnetic
moments  can be described using the following equation:
\begin{equation}\label{eq:MR}
MR= -P_{1}^{2}\ln (1+P_{2}^{2}B^{2})+\frac{P_{3}^{2}B^{2}}{1+P_{4}^{2}B^{2}}.
\end{equation}
The free parameters $P_1$ and $P_2$ depend on several factors such
as a spin scattering amplitude, the exchange integral, the density
of states at the Fermi energy, the spin of the localized magnetic
moments and the average magnetization square~\cite{TOYO,K-F}. The
parameters $P_3$ and $P_4$ depend on the conductivity and the
carrier mobility. In general, Eq.~(\ref{eq:MR}) contains two
competitive terms. The first term describes the negative
contribution due to a spin-dependent scattering mechanism
(\textit{s}-\textit{d} usually in \textit{d}-band ferromagnets,
\textit{s}-\textit{p} in \textit{p}-band ferromagnets~\cite{VOL}).
The second term describes a positive $MR$ due to the usual Lorentz
force contribution. Using this equation we have fitted our
experimental results. They are shown as continuous lines in the
inset of \figurename~\ref{fig:fig5} and the corresponding parameters
are listed in Table~\ref{tab:Tab2}. The fits describe very well
the experimental results at all measured temperatures, indicating
the existence of spin dependent scattering in our GLM, however,
this is not an evidence of magnetic order. Similar
results were already observed in magnetic samples where the
magnetic order was triggered by defects, as in proton irradiated
graphite~\cite{jems08}, proton irradiated ZnO:Li
microwires~\cite{lorjp15}, as well as in samples with magnetic
elements like in single multi-walled carbon nanotubes filled with
Fe nanorods~\cite{JB-MWFE}.

\begin{table}[t]
\begin{tabularx}{\columnwidth}{XXXXX}
  \hline\hline
  $T(K)$&$P_1$ &$P_2({1/\rm K})$&$P_3({1/\rm K})$& $P_4({1/\rm K})$    \\
  \hline
  2   &  0.1643      &  0.0300     & 0.02053        &   0.16316  \vspace{0.05cm}\\
    5   &  0.013       &  1.68       & 0.00936        &   0.05081  \vspace{0.05cm}\\
    10  &  0.0114      &  1.6428     & 0.00541        &   0.00133  \vspace{0.05cm}\\
    15  &  0.0148      &  1.00675    & 0.0015         &   4.075E-7 \vspace{0.05cm}\\
    20  &  0.017       &  0.668      & 1E-4           &   3.07E-7  \vspace{0.05cm}\\

  \hline\hline
\end{tabularx}
\caption{Summary of parameters obtained after fitting the $MR$ using Eq.~(\ref{eq:MR}). }
\label{tab:Tab2}
\end{table}

The magnetic scattering contribution in our
samples can be explained as a consequence of defects present in
the disordered graphite structure of the microwire. Because of the
preparation method used to produce the microwire, we can rule out
the presence of magnetic impurities. The almost zero MR for
temperatures $T \gtrsim 200$~K opens the possibility
to use GLMs in a device under application of high magnetic fields.

\section{Conclusion}
In this work we have prepared a graphitized wire of $1350~\upmu$m
length and lateral area of $\approx 15~\upmu{\rm m}\times130~{\rm
nm}$ inside a diamond matrix by means of ${\rm He}^{+}$
irradiation and  heat treatment. After annealing, the resistivity of the
microwire is only one order of magnitude larger compared to graphite. Also,
% resistance reduces remarkably and
its temperature dependence can be well described by the parallel
contribution of VRH and a metallic-like conduction, similar has been
observed in other carbon based materials. The measured
magnetoresistance can be well described by a semi empirical model,
which takes into account a spin dependent transport mechanism used to
describe the $MR$ of magnetic diluted semiconductors as well as
magnetic carbon-based materials, related to the defect-induced
magnetism.

In summary, the low resistivity
($\rho_{ (300~{\rm K})}=7 \ldots 8\times10^{-5}\Omega$m), the small
resistivity ratio ($\rho(2{\rm K})/\rho(315{\rm K}) \approx 1.275$)
and the very small magnetoresistance, make the graphitized wires
inside diamond candidates to be used as circuit elements.  Our
microwire preparation can be taken as basis for the production and
future application of GLMs inside diamonds. For example, a GLM can be
used to produce a magnetic field by a loop inside the diamond and thus
allows, e.g., further manipulation of NV-centers for future
application in quantum computation. Another possibility is the
application in biology. Considering that diamond is a biocompatible
material, the here used method would enable the design and production
of complete electrical circuits that can allow the monitoring of
electric signals \textit{in-situ} the human body.

\acknowledgements We thank F. Bern for technical support in the
magnetoresistance measurements.

%\bibliography{dm}
%\bibliography{jose,magnetic_carbon}

\begin{thebibliography}{39}%
\makeatletter
\providecommand \@ifxundefined [1]{%
 \@ifx{#1\undefined}
}%
\providecommand \@ifnum [1]{%
 \ifnum #1\expandafter \@firstoftwo
 \else \expandafter \@secondoftwo
 \fi
}%
\providecommand \@ifx [1]{%
 \ifx #1\expandafter \@firstoftwo
 \else \expandafter \@secondoftwo
 \fi
}%
\providecommand \natexlab [1]{#1}%
\providecommand \enquote  [1]{``#1''}%
\providecommand \bibnamefont  [1]{#1}%
\providecommand \bibfnamefont [1]{#1}%
\providecommand \citenamefont [1]{#1}%
\providecommand \href@noop [0]{\@secondoftwo}%
\providecommand \href [0]{\begingroup \@sanitize@url \@href}%
\providecommand \@href[1]{\@@startlink{#1}\@@href}%
\providecommand \@@href[1]{\endgroup#1\@@endlink}%
\providecommand \@sanitize@url [0]{\catcode `\\12\catcode
`\$12\catcode
  `\&12\catcode `\#12\catcode `\^12\catcode `\_12\catcode `\%12\relax}%
\providecommand \@@startlink[1]{}%
\providecommand \@@endlink[0]{}%
\providecommand \url  [0]{\begingroup\@sanitize@url \@url }%
\providecommand \@url [1]{\endgroup\@href {#1}{\urlprefix }}%
\providecommand \urlprefix  [0]{URL }%
\providecommand \Eprint [0]{\href }%
\providecommand \doibase [0]{http://dx.doi.org/}%
\providecommand \selectlanguage [0]{\@gobble}%
\providecommand \bibinfo  [0]{\@secondoftwo}%
\providecommand \bibfield  [0]{\@secondoftwo}%
\providecommand \translation [1]{[#1]}%
\providecommand \BibitemOpen [0]{}%
\providecommand \bibitemStop [0]{}%
\providecommand \bibitemNoStop [0]{.\EOS\space}%
\providecommand \EOS [0]{\spacefactor3000\relax}%
\providecommand \BibitemShut  [1]{\csname bibitem#1\endcsname}%
\let\auto@bib@innerbib\@empty
%</preamble>
\bibitem [{\citenamefont {Vavilov}\ \emph {et~al.}(1974)\citenamefont
  {Vavilov}, \citenamefont {Krasnopevtsev}, \citenamefont {Miljutin},
  \citenamefont {Gorodestsky},\ and\ \citenamefont {Zakharov}}]{VAV}%
  \BibitemOpen
  \bibfield  {author} {\bibinfo {author} {\bibfnamefont {V.~S.}\ \bibnamefont
  {Vavilov}}, \bibinfo {author} {\bibfnamefont {V.~V.}\ \bibnamefont
  {Krasnopevtsev}}, \bibinfo {author} {\bibfnamefont {Y.~V.}\ \bibnamefont
  {Miljutin}}, \bibinfo {author} {\bibfnamefont {A.~E.}\ \bibnamefont
  {Gorodestsky}}, \ and\ \bibinfo {author} {\bibfnamefont {A.~P.}\ \bibnamefont
  {Zakharov}},\ }\bibfield  {title} {\enquote {\bibinfo {title} {On structural
  transitions in ion implanted diamond},}\ }\href@noop {} {\bibfield  {journal}
  {\bibinfo  {journal} {Radiation Effects}\ }\textbf {\bibinfo {volume}
  {22:2}},\ \bibinfo {pages} {141--143} (\bibinfo {year} {1974})}\BibitemShut
  {NoStop}%
\bibitem [{\citenamefont {Hauser}\ \emph {et~al.}(1977)\citenamefont {Hauser},
  \citenamefont {Patel},\ and\ \citenamefont {Rodgers}}]{HAUS1}%
  \BibitemOpen
  \bibfield  {author} {\bibinfo {author} {\bibfnamefont {J.J.}\ \bibnamefont
  {Hauser}}, \bibinfo {author} {\bibfnamefont {J.R.}\ \bibnamefont {Patel}}, \
  and\ \bibinfo {author} {\bibfnamefont {J.W.}\ \bibnamefont {Rodgers}},\
  }\bibfield  {title} {\enquote {\bibinfo {title} {Hard conducting implanted
  diamond layers},}\ }\href@noop {} {\bibfield  {journal} {\bibinfo  {journal}
  {Appl. Phys. Lett.}\ }\textbf {\bibinfo {volume} {30}},\ \bibinfo {pages}
  {129--130} (\bibinfo {year} {1977})}\BibitemShut {NoStop}%
\bibitem [{\citenamefont {Hauser}\ and\ \citenamefont {Patel}(1976)}]{HAUS2}%
  \BibitemOpen
  \bibfield  {author} {\bibinfo {author} {\bibfnamefont {J.J.}\ \bibnamefont
  {Hauser}}\ and\ \bibinfo {author} {\bibfnamefont {J.R.}\ \bibnamefont
  {Patel}},\ }\bibfield  {title} {\enquote {\bibinfo {title} {Hopping
  conductivity in c-implanted amorphous diamond, or how to ruin a perfectly
  good diamond},}\ }\href@noop {} {\bibfield  {journal} {\bibinfo  {journal}
  {Solid State Commun.}\ }\textbf {\bibinfo {volume} {18}},\ \bibinfo {pages}
  {789--790} (\bibinfo {year} {1976})}\BibitemShut {NoStop}%
\bibitem [{\citenamefont {Picollo}\ \emph {et~al.}(2012)\citenamefont
  {Picollo}, \citenamefont {Monticone}, \citenamefont {Olivero}, \citenamefont
  {Fairchild}, \citenamefont {Rubanov}, \citenamefont {Prawer},\ and\
  \citenamefont {Vittone}}]{pic12}%
  \BibitemOpen
  \bibfield  {author} {\bibinfo {author} {\bibfnamefont {F.}~\bibnamefont
  {Picollo}}, \bibinfo {author} {\bibfnamefont {D.~Gatto}\ \bibnamefont
  {Monticone}}, \bibinfo {author} {\bibfnamefont {P.}~\bibnamefont {Olivero}},
  \bibinfo {author} {\bibfnamefont {B.A.}\ \bibnamefont {Fairchild}}, \bibinfo
  {author} {\bibfnamefont {S.}~\bibnamefont {Rubanov}}, \bibinfo {author}
  {\bibfnamefont {S.}~\bibnamefont {Prawer}}, \ and\ \bibinfo {author}
  {\bibfnamefont {E.}~\bibnamefont {Vittone}},\ }\bibfield  {title} {\enquote
  {\bibinfo {title} {Fabrication and electrical characterization of
  three-dimensional graphitic microchannels in single crystal diamond},}\
  }\href@noop {} {\bibfield  {journal} {\bibinfo  {journal} {New Journal of
  Physics}\ }\textbf {\bibinfo {volume} {14}},\ \bibinfo {pages} {053011}
  (\bibinfo {year} {2012})}\BibitemShut {NoStop}%
\bibitem [{\citenamefont {Talapatra}\ \emph {et~al.}(2005)\citenamefont
  {Talapatra}, \citenamefont {Ganesan}, \citenamefont {Kim}, \citenamefont
  {Vajtai}, \citenamefont {Huang}, \citenamefont {Shima}, \citenamefont
  {Ramanath}, \citenamefont {Srivastava}, \citenamefont {Deevi},\ and\
  \citenamefont {Ajayan}}]{TALAP}%
  \BibitemOpen
  \bibfield  {author} {\bibinfo {author} {\bibfnamefont {S.}~\bibnamefont
  {Talapatra}}, \bibinfo {author} {\bibfnamefont {P.~G.}\ \bibnamefont
  {Ganesan}}, \bibinfo {author} {\bibfnamefont {T.}~\bibnamefont {Kim}},
  \bibinfo {author} {\bibfnamefont {R.}~\bibnamefont {Vajtai}}, \bibinfo
  {author} {\bibfnamefont {M.}~\bibnamefont {Huang}}, \bibinfo {author}
  {\bibfnamefont {M.}~\bibnamefont {Shima}}, \bibinfo {author} {\bibfnamefont
  {G.}~\bibnamefont {Ramanath}}, \bibinfo {author} {\bibfnamefont
  {D.}~\bibnamefont {Srivastava}}, \bibinfo {author} {\bibfnamefont {S.~C.}\
  \bibnamefont {Deevi}}, \ and\ \bibinfo {author} {\bibfnamefont {P.~M.}\
  \bibnamefont {Ajayan}},\ }\bibfield  {title} {\enquote {\bibinfo {title}
  {Magnetic properties of point defects in proton irradiated diamond},}\
  }\href@noop {} {\bibfield  {journal} {\bibinfo  {journal} {Phys. Rev. Lett.}\
  }\textbf {\bibinfo {volume} {95}},\ \bibinfo {pages} {097201} (\bibinfo
  {year} {2005})}\BibitemShut {NoStop}%
\bibitem [{\citenamefont {Makgato}\ \emph {et~al.}()\citenamefont {Makgato},
  \citenamefont {Sideras-Haddad}, \citenamefont {Shrivastava}, \citenamefont
  {Madhuku}, \citenamefont {Sekonya}, \citenamefont {Pineda-Vargas},\ and\
  \citenamefont {Joubert}}]{mak14}%
  \BibitemOpen
  \bibfield  {author} {\bibinfo {author} {\bibfnamefont {T.~N.}\ \bibnamefont
  {Makgato}}, \bibinfo {author} {\bibfnamefont {E.}~\bibnamefont
  {Sideras-Haddad}}, \bibinfo {author} {\bibfnamefont {S.}~\bibnamefont
  {Shrivastava}}, \bibinfo {author} {\bibfnamefont {M.}~\bibnamefont
  {Madhuku}}, \bibinfo {author} {\bibfnamefont {K.}~\bibnamefont {Sekonya}},
  \bibinfo {author} {\bibfnamefont {A.}~\bibnamefont {Pineda-Vargas}}, \ and\
  \bibinfo {author} {\bibfnamefont {D.}~\bibnamefont {Joubert}},\ }\href@noop
  {} {\enquote {\bibinfo {title} {Magnetization effects in proton
  micro-irradiated diamond},}\ }\bibinfo {note} {{\rm
  arXiv:1409.2997}}\BibitemShut {NoStop}%
\bibitem [{\citenamefont {Makgato}\ \emph {et~al.}(2016)\citenamefont
  {Makgato}, \citenamefont {Sideras-Haddad}, \citenamefont {Ramos},
  \citenamefont {Garcia-Hernandez}, \citenamefont {Climent-Font}, \citenamefont
  {Munoz-Martin}, \citenamefont {Zucchiatti}, \citenamefont {Shrivastava},\
  and\ \citenamefont {Erasmus}}]{mak16}%
  \BibitemOpen
  \bibfield  {author} {\bibinfo {author} {\bibfnamefont {T.~N.}\ \bibnamefont
  {Makgato}}, \bibinfo {author} {\bibfnamefont {E.}~\bibnamefont
  {Sideras-Haddad}}, \bibinfo {author} {\bibfnamefont {M.~A.}\ \bibnamefont
  {Ramos}}, \bibinfo {author} {\bibfnamefont {M.}~\bibnamefont
  {Garcia-Hernandez}}, \bibinfo {author} {\bibfnamefont {A.}~\bibnamefont
  {Climent-Font}}, \bibinfo {author} {\bibfnamefont {M.}~\bibnamefont
  {Munoz-Martin}}, \bibinfo {author} {\bibfnamefont {A.}~\bibnamefont
  {Zucchiatti}}, \bibinfo {author} {\bibfnamefont {S.}~\bibnamefont
  {Shrivastava}}, \ and\ \bibinfo {author} {\bibfnamefont {R.}~\bibnamefont
  {Erasmus}},\ }\bibfield  {title} {\enquote {\bibinfo {title} {Magnetic
  properties of point defects in proton irradiated diamond},}\ }\href@noop {}
  {\bibfield  {journal} {\bibinfo  {journal} {J. Magn. Magn. Mat.}\ }\textbf
  {\bibinfo {volume} {413}},\ \bibinfo {pages} {76--80} (\bibinfo {year}
  {2016})}\BibitemShut {NoStop}%
\bibitem [{\citenamefont {Esquinazi}\ \emph {et~al.}(2013)\citenamefont
  {Esquinazi}, \citenamefont {Hergert}, \citenamefont {Spemann}, \citenamefont
  {Setzer},\ and\ \citenamefont {Ernst}}]{dim13}%
  \BibitemOpen
  \bibfield  {author} {\bibinfo {author} {\bibfnamefont {P.}~\bibnamefont
  {Esquinazi}}, \bibinfo {author} {\bibfnamefont {W.}~\bibnamefont {Hergert}},
  \bibinfo {author} {\bibfnamefont {D.}~\bibnamefont {Spemann}}, \bibinfo
  {author} {\bibfnamefont {A.}~\bibnamefont {Setzer}}, \ and\ \bibinfo {author}
  {\bibfnamefont {A.}~\bibnamefont {Ernst}},\ }\bibfield  {title} {\enquote
  {\bibinfo {title} {Defect-induced magnetism in solids},}\ }\href@noop {}
  {\bibfield  {journal} {\bibinfo  {journal} {Magnetics, IEEE Transactions on}\
  }\textbf {\bibinfo {volume} {49}},\ \bibinfo {pages} {4668--4674} (\bibinfo
  {year} {2013})}\BibitemShut {NoStop}%
\bibitem [{\citenamefont {Spemann}\ \emph {et~al.}(2001)\citenamefont
  {Spemann}, \citenamefont {Reinert}, \citenamefont {Vogt}, \citenamefont
  {Butz}, \citenamefont {Otte},\ and\ \citenamefont {Zimmer}}]{SPEM}%
  \BibitemOpen
  \bibfield  {author} {\bibinfo {author} {\bibfnamefont {D.}~\bibnamefont
  {Spemann}}, \bibinfo {author} {\bibfnamefont {T.}~\bibnamefont {Reinert}},
  \bibinfo {author} {\bibfnamefont {J.}~\bibnamefont {Vogt}}, \bibinfo {author}
  {\bibfnamefont {T.}~\bibnamefont {Butz}}, \bibinfo {author} {\bibfnamefont
  {K.}~\bibnamefont {Otte}}, \ and\ \bibinfo {author} {\bibfnamefont
  {K.}~\bibnamefont {Zimmer}},\ }\bibfield  {title} {\enquote {\bibinfo {title}
  {Novel test sample for submicron ion-beam analysis},}\ }\href@noop {}
  {\bibfield  {journal} {\bibinfo  {journal} {Nucl. Instr. and Meth. B}\
  }\textbf {\bibinfo {volume} {181}},\ \bibinfo {pages} {186--192} (\bibinfo
  {year} {2001})}\BibitemShut {NoStop}%
\bibitem [{\citenamefont {Mous}\ \emph {et~al.}(1997)\citenamefont {Mous},
  \citenamefont {Haitsma}, \citenamefont {Butz}, \citenamefont {Flagmeyer},
  \citenamefont {Lehmann},\ and\ \citenamefont {Vogt}}]{MOUS}%
  \BibitemOpen
  \bibfield  {author} {\bibinfo {author} {\bibfnamefont {D.~J.~W.}\
  \bibnamefont {Mous}}, \bibinfo {author} {\bibfnamefont {R.~G.}\ \bibnamefont
  {Haitsma}}, \bibinfo {author} {\bibfnamefont {T.}~\bibnamefont {Butz}},
  \bibinfo {author} {\bibfnamefont {R.-H.}\ \bibnamefont {Flagmeyer}}, \bibinfo
  {author} {\bibfnamefont {D.}~\bibnamefont {Lehmann}}, \ and\ \bibinfo
  {author} {\bibfnamefont {J.}~\bibnamefont {Vogt}},\ }\bibfield  {title}
  {\enquote {\bibinfo {title} {The novel ultrastable hvee 3.5 mv
  singletron$^{\rm tm}$ accelerator for nanoprobe applications},}\ }\href@noop
  {} {\bibfield  {journal} {\bibinfo  {journal} {Nucl. Instr. and Meth. B}\
  }\textbf {\bibinfo {volume} {130}},\ \bibinfo {pages} {31--36} (\bibinfo
  {year} {1997})}\BibitemShut {NoStop}%
\bibitem [{\citenamefont {Uzan-Saguy}\ \emph {et~al.}(1995)\citenamefont
  {Uzan-Saguy}, \citenamefont {Cytermann}, \citenamefont {Brener},
  \citenamefont {Richter}, \citenamefont {Shaanan},\ and\ \citenamefont
  {Kalish}}]{UZAN}%
  \BibitemOpen
  \bibfield  {author} {\bibinfo {author} {\bibfnamefont {C.}~\bibnamefont
  {Uzan-Saguy}}, \bibinfo {author} {\bibfnamefont {C.}~\bibnamefont
  {Cytermann}}, \bibinfo {author} {\bibfnamefont {R.}~\bibnamefont {Brener}},
  \bibinfo {author} {\bibfnamefont {V.}~\bibnamefont {Richter}}, \bibinfo
  {author} {\bibfnamefont {M.}~\bibnamefont {Shaanan}}, \ and\ \bibinfo
  {author} {\bibfnamefont {R.}~\bibnamefont {Kalish}},\ }\bibfield  {title}
  {\enquote {\bibinfo {title} {Damage threshold for ion-beam induced
  graphitization of diamond},}\ }\href@noop {} {\bibfield  {journal} {\bibinfo
  {journal} {Appl. Phys. Lett.}\ }\textbf {\bibinfo {volume} {67}},\ \bibinfo
  {pages} {1194--1196} (\bibinfo {year} {1995})}\BibitemShut {NoStop}%
\bibitem [{\citenamefont {Kalish}\ \emph {et~al.}(1999)\citenamefont {Kalish},
  \citenamefont {Reznik}, \citenamefont {Prawer}, \citenamefont {Saada},\ and\
  \citenamefont {Adler}}]{KAL}%
  \BibitemOpen
  \bibfield  {author} {\bibinfo {author} {\bibfnamefont {R.}~\bibnamefont
  {Kalish}}, \bibinfo {author} {\bibfnamefont {A.}~\bibnamefont {Reznik}},
  \bibinfo {author} {\bibfnamefont {S.}~\bibnamefont {Prawer}}, \bibinfo
  {author} {\bibfnamefont {D.}~\bibnamefont {Saada}}, \ and\ \bibinfo {author}
  {\bibfnamefont {J.}~\bibnamefont {Adler}},\ }\bibfield  {title} {\enquote
  {\bibinfo {title} {Ion-implantation-induced defects in diamond and their
  annealing: experiment and simulation},}\ }\href@noop {} {\bibfield  {journal}
  {\bibinfo  {journal} {Physica Status Solidi A-Applied Research}\ }\textbf
  {\bibinfo {volume} {174}},\ \bibinfo {pages} {83--89} (\bibinfo {year}
  {1999})}\BibitemShut {NoStop}%
\bibitem [{\citenamefont {Ziegler}(2013)}]{SRIM}%
  \BibitemOpen
  \bibfield  {author} {\bibinfo {author} {\bibfnamefont {J.}~\bibnamefont
  {Ziegler}},\ }\bibfield  {title} {\enquote {\bibinfo {title} {The stopping
  and range of ions in matter},}\ }\href@noop {} {\bibfield  {journal}
  {\bibinfo  {journal} {www.srim.org}\ } (\bibinfo {year} {2013})}\BibitemShut
  {NoStop}%
\bibitem [{\citenamefont {Robinson}\ and\ \citenamefont
  {Torrens}(1974)}]{ROBIN}%
  \BibitemOpen
  \bibfield  {author} {\bibinfo {author} {\bibfnamefont {M.~T.}\ \bibnamefont
  {Robinson}}\ and\ \bibinfo {author} {\bibfnamefont {I.~M.}\ \bibnamefont
  {Torrens}},\ }\bibfield  {title} {\enquote {\bibinfo {title} {Computer
  simulation of atomic-displacement cascades in solids in the binary-collision
  approximation},}\ }\href@noop {} {\bibfield  {journal} {\bibinfo  {journal}
  {Phys. Rev. B}\ }\textbf {\bibinfo {volume} {9}},\ \bibinfo {pages}
  {508--524} (\bibinfo {year} {1974})}\BibitemShut {NoStop}%
\bibitem [{\citenamefont {Saada}\ \emph {et~al.}(1998)\citenamefont {Saada},
  \citenamefont {Adler},\ and\ \citenamefont {Kalish}}]{SAAD}%
  \BibitemOpen
  \bibfield  {author} {\bibinfo {author} {\bibfnamefont {D.}~\bibnamefont
  {Saada}}, \bibinfo {author} {\bibfnamefont {J.}~\bibnamefont {Adler}}, \ and\
  \bibinfo {author} {\bibfnamefont {R.}~\bibnamefont {Kalish}},\ }\bibfield
  {title} {\enquote {\bibinfo {title} {Transformation of diamond (sp$^3$) to
  graphite (sp$^2$) bonds by ion-impact},}\ }\href@noop {} {\bibfield
  {journal} {\bibinfo  {journal} {Int. J. Mod. Phys. C}\ }\textbf {\bibinfo
  {volume} {09}},\ \bibinfo {pages} {61} (\bibinfo {year} {1998})}\BibitemShut
  {NoStop}%
\bibitem [{\citenamefont {Ma}\ \emph {et~al.}(2014)\citenamefont {Ma},
  \citenamefont {Wu}, \citenamefont {Shen}, \citenamefont {Yan}, \citenamefont
  {Pan},\ and\ \citenamefont {Wang}}]{MA}%
  \BibitemOpen
  \bibfield  {author} {\bibinfo {author} {\bibfnamefont {Z.}~\bibnamefont
  {Ma}}, \bibinfo {author} {\bibfnamefont {J.}~\bibnamefont {Wu}}, \bibinfo
  {author} {\bibfnamefont {W.}~\bibnamefont {Shen}}, \bibinfo {author}
  {\bibfnamefont {L.}~\bibnamefont {Yan}}, \bibinfo {author} {\bibfnamefont
  {X.}~\bibnamefont {Pan}}, \ and\ \bibinfo {author} {\bibfnamefont
  {J.}~\bibnamefont {Wang}},\ }\bibfield  {title} {\enquote {\bibinfo {title}
  {Etching of {CVD} diamond films using oxygen ions in {ECR} plasma},}\
  }\href@noop {} {\bibfield  {journal} {\bibinfo  {journal} {Appl. Surf. Sci.}\
  }\textbf {\bibinfo {volume} {289}},\ \bibinfo {pages} {533--537} (\bibinfo
  {year} {2014})}\BibitemShut {NoStop}%
\bibitem [{\citenamefont {Cancado}\ \emph {et~al.}(2006)\citenamefont
  {Cancado}, \citenamefont {Takai}, \citenamefont {Enoki}, \citenamefont
  {Endo}, \citenamefont {Kim}, \citenamefont {Mizusaki}, \citenamefont {Jorio},
  \citenamefont {Coelho}, \citenamefont {Magalhaes-Paniago},\ and\
  \citenamefont {Pimenta}}]{CANC}%
  \BibitemOpen
  \bibfield  {author} {\bibinfo {author} {\bibfnamefont {L.~G.}\ \bibnamefont
  {Cancado}}, \bibinfo {author} {\bibfnamefont {K.}~\bibnamefont {Takai}},
  \bibinfo {author} {\bibfnamefont {T.}~\bibnamefont {Enoki}}, \bibinfo
  {author} {\bibfnamefont {M.}~\bibnamefont {Endo}}, \bibinfo {author}
  {\bibfnamefont {Y.~A.}\ \bibnamefont {Kim}}, \bibinfo {author} {\bibfnamefont
  {H.}~\bibnamefont {Mizusaki}}, \bibinfo {author} {\bibfnamefont
  {A.}~\bibnamefont {Jorio}}, \bibinfo {author} {\bibfnamefont {L.~N.}\
  \bibnamefont {Coelho}}, \bibinfo {author} {\bibfnamefont {R.}~\bibnamefont
  {Magalhaes-Paniago}}, \ and\ \bibinfo {author} {\bibfnamefont {M.~A.}\
  \bibnamefont {Pimenta}},\ }\bibfield  {title} {\enquote {\bibinfo {title}
  {General equation for the determination of the crystallite size ${L}_a$ of
  nanographite by raman spectroscopy},}\ }\href@noop {} {\bibfield  {journal}
  {\bibinfo  {journal} {Appl. Phys. Lett.}\ }\textbf {\bibinfo {volume} {88}},\
  \bibinfo {pages} {163106--3} (\bibinfo {year} {2006})}\BibitemShut {NoStop}%
\bibitem [{\citenamefont {Rubanov}\ \emph {et~al.}(2015)\citenamefont
  {Rubanov}, \citenamefont {Fairchild}, \citenamefont {Suvorova}, \citenamefont
  {Olivero},\ and\ \citenamefont {Prawer}}]{RUBA}%
  \BibitemOpen
  \bibfield  {author} {\bibinfo {author} {\bibfnamefont {S.}~\bibnamefont
  {Rubanov}}, \bibinfo {author} {\bibfnamefont {B.~A.}\ \bibnamefont
  {Fairchild}}, \bibinfo {author} {\bibfnamefont {A.}~\bibnamefont {Suvorova}},
  \bibinfo {author} {\bibfnamefont {P.}~\bibnamefont {Olivero}}, \ and\
  \bibinfo {author} {\bibfnamefont {S.}~\bibnamefont {Prawer}},\ }\bibfield
  {title} {\enquote {\bibinfo {title} {Structural transformation of implanted
  diamond layers during high temperature annealing},}\ }\href@noop {}
  {\bibfield  {journal} {\bibinfo  {journal} {Nucl. Instr. and Meth. B}\
  }\textbf {\bibinfo {volume} {365}},\ \bibinfo {pages} {50–54} (\bibinfo
  {year} {2015})}\BibitemShut {NoStop}%
\bibitem [{\citenamefont {Barzola-Quiquia}\ \emph {et~al.}(2015)\citenamefont
  {Barzola-Quiquia}, \citenamefont {Esquinazi}, \citenamefont {Lindel},
  \citenamefont {Spemann}, \citenamefont {Muallem},\ and\ \citenamefont
  {Nessim}}]{JBQMWCNT}%
  \BibitemOpen
  \bibfield  {author} {\bibinfo {author} {\bibfnamefont {J.}~\bibnamefont
  {Barzola-Quiquia}}, \bibinfo {author} {\bibfnamefont {P.}~\bibnamefont
  {Esquinazi}}, \bibinfo {author} {\bibfnamefont {M.}~\bibnamefont {Lindel}},
  \bibinfo {author} {\bibfnamefont {D.}~\bibnamefont {Spemann}}, \bibinfo
  {author} {\bibfnamefont {M.}~\bibnamefont {Muallem}}, \ and\ \bibinfo
  {author} {\bibfnamefont {G.D.}\ \bibnamefont {Nessim}},\ }\bibfield  {title}
  {\enquote {\bibinfo {title} {Magnetic order and superconductivity observed in
  bundles of double-wall carbon nanotubes},}\ }\href@noop {} {\bibfield
  {journal} {\bibinfo  {journal} {Carbon}\ }\textbf {\bibinfo {volume} {88}},\
  \bibinfo {pages} {16--25} (\bibinfo {year} {2015})}\BibitemShut {NoStop}%
\bibitem [{\citenamefont {Cholula-Diaz}\ \emph {et~al.}(2014)\citenamefont
  {Cholula-Diaz}, \citenamefont {Barzola-Quiquia}, \citenamefont {Krautscheid},
  \citenamefont {Teschner},\ and\ \citenamefont {Esquinazi}}]{CHOL}%
  \BibitemOpen
  \bibfield  {author} {\bibinfo {author} {\bibfnamefont {J.~L.}\ \bibnamefont
  {Cholula-Diaz}}, \bibinfo {author} {\bibfnamefont {J.}~\bibnamefont
  {Barzola-Quiquia}}, \bibinfo {author} {\bibfnamefont {H.}~\bibnamefont
  {Krautscheid}}, \bibinfo {author} {\bibfnamefont {U.}~\bibnamefont
  {Teschner}}, \ and\ \bibinfo {author} {\bibfnamefont {P.}~\bibnamefont
  {Esquinazi}},\ }\bibfield  {title} {\enquote {\bibinfo {title} {Synthesis and
  magnetotransport properties of nanocrystalline graphite prepared by aerosol
  assisted chemical vapor deposition},}\ }\href@noop {} {\bibfield  {journal}
  {\bibinfo  {journal} {Carbon}\ }\textbf {\bibinfo {volume} {67}},\ \bibinfo
  {pages} {10--16} (\bibinfo {year} {2014})}\BibitemShut {NoStop}%
\bibitem [{\citenamefont {Barzola-Quiquia}\ \emph {et~al.}(2008)\citenamefont
  {Barzola-Quiquia}, \citenamefont {Yao}, \citenamefont {R\"odiger},
  \citenamefont {Schindler},\ and\ \citenamefont {Esquinazi}}]{JB1}%
  \BibitemOpen
  \bibfield  {author} {\bibinfo {author} {\bibfnamefont {J.}~\bibnamefont
  {Barzola-Quiquia}}, \bibinfo {author} {\bibfnamefont {J.-L.}\ \bibnamefont
  {Yao}}, \bibinfo {author} {\bibfnamefont {P.}~\bibnamefont {R\"odiger}},
  \bibinfo {author} {\bibfnamefont {K.}~\bibnamefont {Schindler}}, \ and\
  \bibinfo {author} {\bibfnamefont {P.}~\bibnamefont {Esquinazi}},\ }\bibfield
  {title} {\enquote {\bibinfo {title} {Sample size effects on the transport
  characteristics of mesoscopic graphite samples},}\ }\href@noop {} {\bibfield
  {journal} {\bibinfo  {journal} {Phys. Stat. Sol. (a)}\ }\textbf {\bibinfo
  {volume} {205}},\ \bibinfo {pages} {2924--2933} (\bibinfo {year}
  {2008})}\BibitemShut {NoStop}%
\bibitem [{\citenamefont {Zoraghi}\ \emph {et~al.}(2017)\citenamefont
  {Zoraghi}, \citenamefont {Barzola-Quiquia}, \citenamefont {Stiller},
  \citenamefont {Setzer}, \citenamefont {Esquinazi}, \citenamefont {Kloess},
  \citenamefont {Muenster}, \citenamefont {L\"uhmann},\ and\ \citenamefont
  {Estrela-Lopis}}]{MZ1}%
  \BibitemOpen
  \bibfield  {author} {\bibinfo {author} {\bibfnamefont {M.}~\bibnamefont
  {Zoraghi}}, \bibinfo {author} {\bibfnamefont {J.}~\bibnamefont
  {Barzola-Quiquia}}, \bibinfo {author} {\bibfnamefont {M.}~\bibnamefont
  {Stiller}}, \bibinfo {author} {\bibfnamefont {A.}~\bibnamefont {Setzer}},
  \bibinfo {author} {\bibfnamefont {P.}~\bibnamefont {Esquinazi}}, \bibinfo
  {author} {\bibfnamefont {G.~H.}\ \bibnamefont {Kloess}}, \bibinfo {author}
  {\bibfnamefont {T.}~\bibnamefont {Muenster}}, \bibinfo {author}
  {\bibfnamefont {T.}~\bibnamefont {L\"uhmann}}, \ and\ \bibinfo {author}
  {\bibfnamefont {I.}~\bibnamefont {Estrela-Lopis}},\ }\bibfield  {title}
  {\enquote {\bibinfo {title} {Influence of rhombohedral stacking order in the
  electrical resistance of bulk and mesoscopic graphite},}\ }\href@noop {}
  {\bibfield  {journal} {\bibinfo  {journal} {Phys. Rev. B}\ }\textbf {\bibinfo
  {volume} {95}},\ \bibinfo {pages} {045308} (\bibinfo {year}
  {2017})}\BibitemShut {NoStop}%
\bibitem [{\citenamefont {Mott}(1968)}]{MOTT}%
  \BibitemOpen
  \bibfield  {author} {\bibinfo {author} {\bibfnamefont {N.~F.}\ \bibnamefont
  {Mott}},\ }\bibfield  {title} {\enquote {\bibinfo {title} {Conduction in
  glasses containing transition metal ions},}\ }\href@noop {} {\bibfield
  {journal} {\bibinfo  {journal} {J. Non-Cryst. Solids}\ }\textbf {\bibinfo
  {volume} {1}},\ \bibinfo {pages} {1--17} (\bibinfo {year}
  {1968})}\BibitemShut {NoStop}%
\bibitem [{\citenamefont {Matsubara}\ \emph {et~al.}(1990)\citenamefont
  {Matsubara}, \citenamefont {Sugihara},\ and\ \citenamefont
  {Tsuzuku}}]{MATSU}%
  \BibitemOpen
  \bibfield  {author} {\bibinfo {author} {\bibfnamefont {K.}~\bibnamefont
  {Matsubara}}, \bibinfo {author} {\bibfnamefont {K.}~\bibnamefont {Sugihara}},
  \ and\ \bibinfo {author} {\bibfnamefont {T.}~\bibnamefont {Tsuzuku}},\
  }\bibfield  {title} {\enquote {\bibinfo {title} {Electrical resistance in the
  c direction of graphite},}\ }\href@noop {} {\bibfield  {journal} {\bibinfo
  {journal} {Phys. Rev. B}\ }\textbf {\bibinfo {volume} {41}},\ \bibinfo
  {pages} {969--974} (\bibinfo {year} {1990})}\BibitemShut {NoStop}%
\bibitem [{\citenamefont {Garcia}\ \emph {et~al.}(2012)\citenamefont {Garcia},
  \citenamefont {Esquinazi}, \citenamefont {Barzola-Quiquia},\ and\
  \citenamefont {Dusari}}]{GARC}%
  \BibitemOpen
  \bibfield  {author} {\bibinfo {author} {\bibfnamefont {N.}~\bibnamefont
  {Garcia}}, \bibinfo {author} {\bibfnamefont {P.}~\bibnamefont {Esquinazi}},
  \bibinfo {author} {\bibfnamefont {J.}~\bibnamefont {Barzola-Quiquia}}, \ and\
  \bibinfo {author} {\bibfnamefont {S.}~\bibnamefont {Dusari}},\ }\bibfield
  {title} {\enquote {\bibinfo {title} {Evidence for semiconducting behavior
  with a narrow band gap of {Bernal} graphite},}\ }\href@noop {} {\bibfield
  {journal} {\bibinfo  {journal} {New Journal of Physics}\ }\textbf {\bibinfo
  {volume} {14}},\ \bibinfo {pages} {053015 (14pp)} (\bibinfo {year}
  {2012})}\BibitemShut {NoStop}%
\bibitem [{\citenamefont {L{\"u}hmann}(2015)}]{TL-16}%
  \BibitemOpen
  \bibfield  {author} {\bibinfo {author} {\bibfnamefont {T.}~\bibnamefont
  {L{\"u}hmann}},\ }\emph {\bibinfo {title} {{ 3D Ionenstrahlschreiben in
  Diamant zur Erzeugung von Graphitstrukturen und deren Charakterisierung}}},\
  \href@noop {} {Master's thesis},\ \bibinfo  {school} {University Leipzig},
  \bibinfo {address} {Germany} (\bibinfo {year} {2015})\BibitemShut {NoStop}%
\bibitem [{\citenamefont {Raj}\ and\ \citenamefont {Joy}(2015)}]{RAJ}%
  \BibitemOpen
  \bibfield  {author} {\bibinfo {author} {\bibfnamefont {K.~G.}\ \bibnamefont
  {Raj}}\ and\ \bibinfo {author} {\bibfnamefont {P.~A.}\ \bibnamefont {Joy}},\
  }\bibfield  {title} {\enquote {\bibinfo {title} {Cross over from {3D}
  variable range hopping to the {2D} weak localization conduction mechanism in
  disordered carbon with the extent of graphitization},}\ }\href@noop {}
  {\bibfield  {journal} {\bibinfo  {journal} {Phys. Chem. Chem. Phys.}\
  }\textbf {\bibinfo {volume} {17}},\ \bibinfo {pages} {16178--16185} (\bibinfo
  {year} {2015})}\BibitemShut {NoStop}%
\bibitem [{\citenamefont {Khan}\ \emph {et~al.}(2008)\citenamefont {Khan},
  \citenamefont {Husain}, \citenamefont {Perng}, \citenamefont {Salah},\ and\
  \citenamefont {Habib}}]{KHA}%
  \BibitemOpen
  \bibfield  {author} {\bibinfo {author} {\bibfnamefont {Z.~H.}\ \bibnamefont
  {Khan}}, \bibinfo {author} {\bibfnamefont {M.}~\bibnamefont {Husain}},
  \bibinfo {author} {\bibfnamefont {T.~P.}\ \bibnamefont {Perng}}, \bibinfo
  {author} {\bibfnamefont {N.}~\bibnamefont {Salah}}, \ and\ \bibinfo {author}
  {\bibfnamefont {S.}~\bibnamefont {Habib}},\ }\bibfield  {title} {\enquote
  {\bibinfo {title} {Electrical transport via variable range hopping in an
  individual multi-wall carbon nanotube},}\ }\href@noop {} {\bibfield
  {journal} {\bibinfo  {journal} {J. Phys.: Condens. Matter}\ }\textbf
  {\bibinfo {volume} {20}},\ \bibinfo {pages} {475207 (7pp)} (\bibinfo {year}
  {2008})}\BibitemShut {NoStop}%
\bibitem [{\citenamefont {Olivero}\ \emph {et~al.}(2010)\citenamefont
  {Olivero}, \citenamefont {Amato}, \citenamefont {Bellotti}, \citenamefont
  {Borini}, \citenamefont {Giudice}, \citenamefont {Picollo},\ and\
  \citenamefont {Vittone}}]{OLIV1}%
  \BibitemOpen
  \bibfield  {author} {\bibinfo {author} {\bibfnamefont {P.}~\bibnamefont
  {Olivero}}, \bibinfo {author} {\bibfnamefont {G.}~\bibnamefont {Amato}},
  \bibinfo {author} {\bibfnamefont {F.}~\bibnamefont {Bellotti}}, \bibinfo
  {author} {\bibfnamefont {S.}~\bibnamefont {Borini}}, \bibinfo {author}
  {\bibfnamefont {A.~Lo}\ \bibnamefont {Giudice}}, \bibinfo {author}
  {\bibfnamefont {F.}~\bibnamefont {Picollo}}, \ and\ \bibinfo {author}
  {\bibfnamefont {E.}~\bibnamefont {Vittone}},\ }\bibfield  {title} {\enquote
  {\bibinfo {title} {Direct fabrication and iv characterization of sub-surface
  conductive channels in diamond with mev ion implantation},}\ }\href@noop {}
  {\bibfield  {journal} {\bibinfo  {journal} {Eur. Phys. J. B}\ }\textbf
  {\bibinfo {volume} {75}},\ \bibinfo {pages} {127–132} (\bibinfo {year}
  {2010})}\BibitemShut {NoStop}%
\bibitem [{\citenamefont {Kempa}\ \emph {et~al.}(2002)\citenamefont {Kempa},
  \citenamefont {Esquinazi},\ and\ \citenamefont {Kopelevich}}]{KEMP}%
  \BibitemOpen
  \bibfield  {author} {\bibinfo {author} {\bibfnamefont {H.}~\bibnamefont
  {Kempa}}, \bibinfo {author} {\bibfnamefont {P.}~\bibnamefont {Esquinazi}}, \
  and\ \bibinfo {author} {\bibfnamefont {Y.}~\bibnamefont {Kopelevich}},\
  }\bibfield  {title} {\enquote {\bibinfo {title} {Field-induced
  metal-insulator transition in the c-axis resistivity of graphite},}\
  }\href@noop {} {\bibfield  {journal} {\bibinfo  {journal} {Phys. Rev. B}\
  }\textbf {\bibinfo {volume} {65}},\ \bibinfo {pages} {241101(R)} (\bibinfo
  {year} {2002})}\BibitemShut {NoStop}%
\bibitem [{\citenamefont {Grill}(1999)}]{GRILL}%
  \BibitemOpen
  \bibfield  {author} {\bibinfo {author} {\bibfnamefont {A.}~\bibnamefont
  {Grill}},\ }\bibfield  {title} {\enquote {\bibinfo {title} {Electrical and
  optical properties of diamond-like carbon},}\ }\href@noop {} {\bibfield
  {journal} {\bibinfo  {journal} {Thin Solid Films}\ }\textbf {\bibinfo
  {volume} {355–356}},\ \bibinfo {pages} {189–193} (\bibinfo {year}
  {1999})}\BibitemShut {NoStop}%
\bibitem [{\citenamefont {Hauser}(1975)}]{HAUS3}%
  \BibitemOpen
  \bibfield  {author} {\bibinfo {author} {\bibfnamefont {J.J.}\ \bibnamefont
  {Hauser}},\ }\bibfield  {title} {\enquote {\bibinfo {title} {Hopping
  conductivity in amorphous carbon films},}\ }\href@noop {} {\bibfield
  {journal} {\bibinfo  {journal} {Solid State Commun.}\ }\textbf {\bibinfo
  {volume} {17}},\ \bibinfo {pages} {1577--1580} (\bibinfo {year}
  {1975})}\BibitemShut {NoStop}%
\bibitem [{\citenamefont {Kopelevich}\ \emph {et~al.}(2003)\citenamefont
  {Kopelevich}, \citenamefont {Torres}, \citenamefont {da~Silva}, \citenamefont
  {Mrowka}, \citenamefont {Kempa},\ and\ \citenamefont {Esquinazi}}]{KOPE1}%
  \BibitemOpen
  \bibfield  {author} {\bibinfo {author} {\bibfnamefont {Y.}~\bibnamefont
  {Kopelevich}}, \bibinfo {author} {\bibfnamefont {J.~H.~S.}\ \bibnamefont
  {Torres}}, \bibinfo {author} {\bibfnamefont {R.~R.}\ \bibnamefont
  {da~Silva}}, \bibinfo {author} {\bibfnamefont {F.}~\bibnamefont {Mrowka}},
  \bibinfo {author} {\bibfnamefont {H.}~\bibnamefont {Kempa}}, \ and\ \bibinfo
  {author} {\bibfnamefont {P.}~\bibnamefont {Esquinazi}},\ }\bibfield  {title}
  {\enquote {\bibinfo {title} {Reentrant metallic behavior of graphite in the
  quantum limit},}\ }\href@noop {} {\bibfield  {journal} {\bibinfo  {journal}
  {Phys. Rev. Lett.}\ }\textbf {\bibinfo {volume} {90}},\ \bibinfo {pages}
  {156402--1--4} (\bibinfo {year} {2003})}\BibitemShut {NoStop}%
\bibitem [{\citenamefont {Toyozawa}(1962)}]{TOYO}%
  \BibitemOpen
  \bibfield  {author} {\bibinfo {author} {\bibfnamefont {Y.}~\bibnamefont
  {Toyozawa}},\ }\bibfield  {title} {\enquote {\bibinfo {title} {Theory of
  localized spins and negative magnetoresistance in the metallic impurity
  conduction},}\ }\href@noop {} {\bibfield  {journal} {\bibinfo  {journal} {J.
  Phys. Soc. Japan}\ }\textbf {\bibinfo {volume} {17}},\ \bibinfo {pages}
  {986--1004} (\bibinfo {year} {1962})}\BibitemShut {NoStop}%
\bibitem [{\citenamefont {Khosla}\ and\ \citenamefont {Fischer}(1970)}]{K-F}%
  \BibitemOpen
  \bibfield  {author} {\bibinfo {author} {\bibfnamefont {R.}~\bibnamefont
  {Khosla}}\ and\ \bibinfo {author} {\bibfnamefont {J.}~\bibnamefont
  {Fischer}},\ }\bibfield  {title} {\enquote {\bibinfo {title}
  {Magnetoresistance in degenerate {CdS}: Localized magnetic moments},}\
  }\href@noop {} {\bibfield  {journal} {\bibinfo  {journal} {Phys. Rev. B}\
  }\textbf {\bibinfo {volume} {2}},\ \bibinfo {pages} {4084--4097} (\bibinfo
  {year} {1970})}\BibitemShut {NoStop}%
\bibitem [{\citenamefont {Volnianska}\ and\ \citenamefont
  {Boguslawski}(2010)}]{VOL}%
  \BibitemOpen
  \bibfield  {author} {\bibinfo {author} {\bibfnamefont {O.}~\bibnamefont
  {Volnianska}}\ and\ \bibinfo {author} {\bibfnamefont {P.}~\bibnamefont
  {Boguslawski}},\ }\bibfield  {title} {\enquote {\bibinfo {title} {Magnetism
  of solids resulting from spin polarization of p orbitals},}\ }\href@noop {}
  {\bibfield  {journal} {\bibinfo  {journal} {J Phys: Condens Matter}\ }\textbf
  {\bibinfo {volume} {22}},\ \bibinfo {pages} {073202 (19pp)} (\bibinfo {year}
  {2010})}\BibitemShut {NoStop}%
\bibitem [{\citenamefont {Esquinazi}\ \emph {et~al.}(2010)\citenamefont
  {Esquinazi}, \citenamefont {Barzola-Quiquia}, \citenamefont {Spemann},
  \citenamefont {Rothermel}, \citenamefont {Ohldag}, \citenamefont {Garc\'ia},
  \citenamefont {Setzer},\ and\ \citenamefont {Butz}}]{jems08}%
  \BibitemOpen
  \bibfield  {author} {\bibinfo {author} {\bibfnamefont {P.}~\bibnamefont
  {Esquinazi}}, \bibinfo {author} {\bibfnamefont {J.}~\bibnamefont
  {Barzola-Quiquia}}, \bibinfo {author} {\bibfnamefont {D.}~\bibnamefont
  {Spemann}}, \bibinfo {author} {\bibfnamefont {M.}~\bibnamefont {Rothermel}},
  \bibinfo {author} {\bibfnamefont {H.}~\bibnamefont {Ohldag}}, \bibinfo
  {author} {\bibfnamefont {N.}~\bibnamefont {Garc\'ia}}, \bibinfo {author}
  {\bibfnamefont {A.}~\bibnamefont {Setzer}}, \ and\ \bibinfo {author}
  {\bibfnamefont {T.}~\bibnamefont {Butz}},\ }\bibfield  {title} {\enquote
  {\bibinfo {title} {Magnetic order in graphite: Experimental evidence,
  intrinsic and extrinsic difficulties},}\ }\href@noop {} {\bibfield  {journal}
  {\bibinfo  {journal} {J. Magn. Magn. Mat.}\ }\textbf {\bibinfo {volume}
  {322}},\ \bibinfo {pages} {1156--1161} (\bibinfo {year} {2010})}\BibitemShut
  {NoStop}%
\bibitem [{\citenamefont {Lorite}\ \emph {et~al.}(2015)\citenamefont {Lorite},
  \citenamefont {Zandalazini}, \citenamefont {Esquinazi}, \citenamefont
  {Spemann}, \citenamefont {Friedl\"ander}, \citenamefont {P\"oppl},
  \citenamefont {Michalsky}, \citenamefont {Grundmann}, \citenamefont {Vogt},
  \citenamefont {Meijer}, \citenamefont {Heluani}, \citenamefont {Ohldag},
  \citenamefont {Adeagbo}, \citenamefont {Nayak}, \citenamefont {Hergert},
  \citenamefont {Ernst},\ and\ \citenamefont {Hoffmann}}]{lorjp15}%
  \BibitemOpen
  \bibfield  {author} {\bibinfo {author} {\bibfnamefont {I.}~\bibnamefont
  {Lorite}}, \bibinfo {author} {\bibfnamefont {C.}~\bibnamefont {Zandalazini}},
  \bibinfo {author} {\bibfnamefont {P.}~\bibnamefont {Esquinazi}}, \bibinfo
  {author} {\bibfnamefont {D.}~\bibnamefont {Spemann}}, \bibinfo {author}
  {\bibfnamefont {S.}~\bibnamefont {Friedl\"ander}}, \bibinfo {author}
  {\bibfnamefont {A.}~\bibnamefont {P\"oppl}}, \bibinfo {author} {\bibfnamefont
  {T.}~\bibnamefont {Michalsky}}, \bibinfo {author} {\bibfnamefont
  {M.}~\bibnamefont {Grundmann}}, \bibinfo {author} {\bibfnamefont
  {J.}~\bibnamefont {Vogt}}, \bibinfo {author} {\bibfnamefont {J.}~\bibnamefont
  {Meijer}}, \bibinfo {author} {\bibfnamefont {S.P.}\ \bibnamefont {Heluani}},
  \bibinfo {author} {\bibfnamefont {H.}~\bibnamefont {Ohldag}}, \bibinfo
  {author} {\bibfnamefont {W.A.}\ \bibnamefont {Adeagbo}}, \bibinfo {author}
  {\bibfnamefont {S.K.}\ \bibnamefont {Nayak}}, \bibinfo {author}
  {\bibfnamefont {W.}~\bibnamefont {Hergert}}, \bibinfo {author} {\bibfnamefont
  {A.}~\bibnamefont {Ernst}}, \ and\ \bibinfo {author} {\bibfnamefont
  {M.}~\bibnamefont {Hoffmann}},\ }\bibfield  {title} {\enquote {\bibinfo
  {title} {Study of the negative magneto-resistance of single proton-implanted
  lithium-doped {ZnO} microwires},}\ }\href@noop {} {\bibfield  {journal}
  {\bibinfo  {journal} {J. Phys.: Condens. Matter}\ }\textbf {\bibinfo {volume}
  {27}},\ \bibinfo {pages} {256002} (\bibinfo {year} {2015})}\BibitemShut
  {NoStop}%
\bibitem [{\citenamefont {Barzola-Quiquia}\ \emph {et~al.}(2012)\citenamefont
  {Barzola-Quiquia}, \citenamefont {Klingner}, \citenamefont {Kr\"uger},
  \citenamefont {Molle}, \citenamefont {Esquinazi}, \citenamefont {Leonhardt},\
  and\ \citenamefont {Martinez}}]{JB-MWFE}%
  \BibitemOpen
  \bibfield  {author} {\bibinfo {author} {\bibfnamefont {J.}~\bibnamefont
  {Barzola-Quiquia}}, \bibinfo {author} {\bibfnamefont {N.}~\bibnamefont
  {Klingner}}, \bibinfo {author} {\bibfnamefont {J.}~\bibnamefont {Kr\"uger}},
  \bibinfo {author} {\bibfnamefont {A.}~\bibnamefont {Molle}}, \bibinfo
  {author} {\bibfnamefont {P.}~\bibnamefont {Esquinazi}}, \bibinfo {author}
  {\bibfnamefont {A.}~\bibnamefont {Leonhardt}}, \ and\ \bibinfo {author}
  {\bibfnamefont {M.~T.}\ \bibnamefont {Martinez}},\ }\bibfield  {title}
  {\enquote {\bibinfo {title} {Quantum oscillations and ferromagnetic
  hysteresis observed in iron filled multiwall carbon nanotubes},}\ }\href@noop
  {} {\bibfield  {journal} {\bibinfo  {journal} {Nanotechnology}\ }\textbf
  {\bibinfo {volume} {23}},\ \bibinfo {pages} {015707 7(pp)} (\bibinfo {year}
  {2012})}\BibitemShut {NoStop}%
\end{thebibliography}

%merlin.mbs apsrev4-1.bst 2010-07-25 4.21a (PWD, AO, DPC) hacked
%Control: key (0)
%Control: author (0) dotless jnrlst
%Control: editor formatted (1) identically to author
%Control: production of article title (0) allowed
%Control: page (1) range
%Control: year (0) verbatim
%Control: production of eprint (0) enabled
%

\end{document}